\documentclass[iop]{emulateapj}
\usepackage{amssymb,graphicx,longtable,multirow}
\usepackage{wasysym}
\usepackage{color}
\usepackage{epsfig}
\usepackage{epstopdf}
\usepackage[]{natbib}

\def\msol{\hbox{$\rm\thinspace M_{\odot}\thinspace$}} 
 
\def\etal{{\it et al.\ }}

\newcommand{\be}{\begin{equation}}
\newcommand{\ba}{\begin{eqnarray}}
\newcommand{\ee}{\end{equation}}
\newcommand{\ea}{\end{eqnarray}}

\newcommand{\cms}{\ensuremath{\mathrm{cm~s}^{-1}}}
\newcommand{\dtx}{\ensuremath{\Delta t_{\rm ex}}}
\newcommand{\kms}{\ensuremath{\mathrm{km~s}^{-1}}}
\newcommand{\pcc}{\ensuremath{\mathrm{cm}^{-3}}}

\newcommand{\nuclei}[2]{\ensuremath{\mathrm{^{#1}#2}}}

\newcommand{\nickel}[1][58]{\nuclei{#1}{Ni}}

\shorttitle{Tidal Tail Mass Ejections}
\shortauthors{Raskin \& Kasen, 2013}

\begin{document}

\title{Tidal Tail Ejection as a Signature of Type~Ia Supernovae from White Dwarf Mergers}
\author{Cody Raskin\altaffilmark{1} \& Daniel Kasen\altaffilmark{1,2}}
\altaffiltext{1}{Nuclear Science Division, Lawrence Berkeley National Laboratory, Berkeley, CA, USA} 
\altaffiltext{2}{Departments of Physics and Astronomy, University of California, Berkeley, CA, USA}

\begin{abstract}
The merger of two white dwarfs may be preceded  by the ejection of some mass in ``tidal tails", creating a circumstellar medium around the system.  We consider the variety of observational signatures from this material, which  depend on the lag time  between the start of the merger  and the ultimate explosion (assuming one occurs) of the system in a Type~Ia supernova. If the time lag is fairly short, the  interaction of the supernova ejecta with the tails could lead to detectable shock emission at radio, optical, and/or x-ray wavelengths.  At somewhat later times, the  tails  produce relatively broad  NaID absorption lines with velocity widths of order the white dwarf escape
speed ($\sim 1000$~\kms). That none of these  signatures have been detected in normal SNe~Ia  constrains the  lag time to be either  very short ($\lesssim 100$~s) or fairly long ($\gtrsim 100$~yr). If the tails have  expanded and cooled over timescales $\sim 10^4$~yr, they could be observable through narrow NaID and CaII H\&K absorption lines in the spectra, which are seen in some fraction of SNe~Ia.  Using a combination of 3D and 1D hydrodynamical codes, we model the mass-loss from tidal interactions in binary systems, and the subsequent interactions with the interstellar medium, which produce a slow-moving, dense shell of  gas. We synthesize NaID  line profiles by ray-casting through this shell, and show that in some circumstances tidal tails could be responsible for  narrow absorptions similar to those observed.
\end{abstract}

\section{Introduction}

Type Ia Supernovae (SNe~Ia) are important cosmological tools due to their remarkably standard light curves \citep{Pskovskii1977,Phillips1993} which lend to their use as standard candles \citep{Colgate1979,Branch1992}. They are thought to be the result of a thermonuclear runaway in  a degenerate white dwarf star, producing large amounts ($\approx 0.6$\msol) of radioactive \nickel[56], however the precise mechanism for producing this detonation remains elusive.

Currently, there are two major candidates for SNeIa progenitor systems. In the single degenerate scenario, the system consists of a white dwarf that accretes mass from a main-sequence or evolved companion star \citep{Whelan1973,Nomoto1982a,Hillebrandt2000}. Under certain constraints on the mass-accretion rate, the white dwarf mass can near the Chandraskehar limit and a thermonuclear runaway will be ignited near the center \citep{Nomoto1982a,Nomoto1982b}. In the double degenerate merger scenario, two white dwarfs in a close binary inspiral and coalesce.  If the merger is violent enough, a detonation could be triggered  on the dynamical timescale ($\sim 10-100$~secs), promptly exploding the system  \citep{Pakmor2010,Pakmor2011,Guillochon2010}. Absent a detonation, the resulting configuration is a more massive, degenerate remnant surrounded by an accretion disk \citep{Yoon2007}.  The disk will evolve viscously on timescales $10^4 - 10^8$ s, heating the shear layers and driving the remnant to a more spherical  state \citep{Shen2012,Schwab2012,Ji2013}. The envelope will then evolve thermally on timescales approaching $10^3-10^4$~years, while the core thermal timescale is $\sim 10^5$~years.  If the secondary white dwarf is composed of helium, stable shell burning may ensue on similarly long timescales.
 
 The final outcome of such a merged remnant is still debated.  Pioneering calculations \citep{Saio2004} indicated that system likely evolved to an off-center, non-violent ignition, with the end result being collapse to a neutron star.   Subsequent studies have  challenged this conclusion, suggesting that the remnant may instead evolve to  central ignition (and a SN~Ia) on a viscous or thermal timescale \citep{Yoon2007, vanKerkwijk2010, Ji2013}.  Other detailed studies of the post-merger evolution, however, suggest that off-center ignition and collapse is indeed the more likely result in most cases \citep{Shen2012,Schwab2012}.   The remnant evolution apparently depends on the binary mass ratio and the initial synchronization conditions \citep{Zhu2012}, and may be sensitive to the  treatment of  e.g.,  mass loss  -- hence,  thermonuclear explosions may be allowed in certain limited regions of parameter space.  In addition, entirely different ways of triggering a detonation in a double degenerate system may be relevant.   \citet{Katz2012} suggest that if the binary is in a triple system, Kozai resonance effects can lead to multiple close passages (with period of $\sim 10^4$~yrs)  culminating in a near head on collision of the stars \citep{Rosswog2009, Raskin2009a, Raskin2010, Hawley2012}.

%As this accretion disk evolves over thermal time-scales, shear forces can initiate a detonation on the surface of the remnant, or the steady disk accretion can simply push the remnant over the Chandrasekhar limit. Though stellar population synthesis models predict many more single degenerate than double degenerate systems (\eg Greggio 2005), the relative frequency of these two mechanisms contributing to the total SNeIa rate is uncertain.

Given the difficulty of the theory, it would be valuable to identify empirical signatures of double degenerate mergers.  One of the more dramatic consequences of the dynamics of compact object mergers is the tidal stripping and ejection of mass which occurs just prior to coalescence.  These ``tidal tail'' ejections are a robust feature of merger calculations and consist of $10^{-4} - 10^{-2}~\msol$ of material \citep{Dan2011} moving at the escape velocity, $\sim 10^8~{\rm cm~s^{-1}}$, and  concentrated in the equatorial regions. Though the total mass expelled is a small fraction of the system mass, it is still  enough to produce a relatively dense surrounding medium.  If the circumstellar material (CSM) is relatively nearby  ($r \lesssim 10^{16}$~cm) when a SN~Ia explodes, the interaction of the supernova ejecta could lead to detectable shock emission at radio, optical, and/or x-ray wavelengths. If the CSM has instead expanded to larger radii and cooled, it may still be observable through narrow absorption lines in the spectra, in particular the resonance doublet lines of NaID and CaII~H\&K.

Recently, spectral observations of a handful of SNeIa have discovered such narrow CSM absorptions at velocities  ($\sim10-100$ \kms)  \citep{Patat2007, Borkowski2009, Simon2009, Sternberg2011, Foley2012}.  In a few cases, the strength of the lines are observed to vary with time, suggesting that the CSM is relatively near to the supernova. In most other cases, the lines do not vary with time, but are observed to be preferentially blue-shifted for SNe~Ia \citep{Sternberg2011}, which  suggests that they are statistically related to mass-loss from the SNeIa progenitor system.  Typically, single degenerate systems are assumed to be the culprit for this CSM, the result of wind mass loss from a red giant or asymptotic giant branch companion, perhaps shaped by novae eruptions from the white dwarf  \citep{Moore2012}.  Recently, \cite{Shen2013} have suggested that for certain classes of double degenerates (C/O + He WDs) nova eruptions in accreted surface layers could produce the necessary absorber material. We show here that the tidal tails from double degenerate mergers, when combined with interactions with the interstellar medium (ISM), can also result in a dense cloud of blue-shifted, absorber material moving at a rage of velocities, assuming that the time-scale between the mass ejection and the SN~Ia explosion is long. 

In this paper, we  explore the signatures of CSM from tidal tails in double degenerate mergers. In \S2, we discuss the combination of 3D and 1D numerical approaches we use for modeling the white dwarf merger as well as the subsequent interaction of ejected material with the ISM. We also compare the results of these numerical studies to analytical estimates of this interaction in \S2.1. In \S3, we estimate the optical depth of the absorbing material and use a spectral synthesis code to produce an absorption profile of the NaID line. Finally, in \S4, we summarize our conclusions and discuss potential avenues for followup. 

\section{Tidal Tail Ejecta}

To determine the mass-loss during a double degenerate white dwarf merger, we use  a 3D smoothed particle hydrodynamics (SPH) code called SNSPH \citep{Fryer2006}. SPH codes are lagrangian by design, and so naturally conserve angular momentum. This feature is crucial for estimating the fraction of the donor star that attains escape velocity.   Following the procedure laid out in \citet{Raskin2012} for generating accurate initial conditions for a white dwarf merger, we simulate a merger pair consisting of a 1.20\msol primary and a 0.64\msol secondary with $5\times10^5$ equal mass particles throughout each star. This mass pair is disparate enough for the secondary to experience strong tidal forces, while still close enough in mass for the merger to exhibit catastrophic disruption of the smaller mass star. 

\begin{figure}[ht]
\centering
\includegraphics[width=0.45\textwidth]{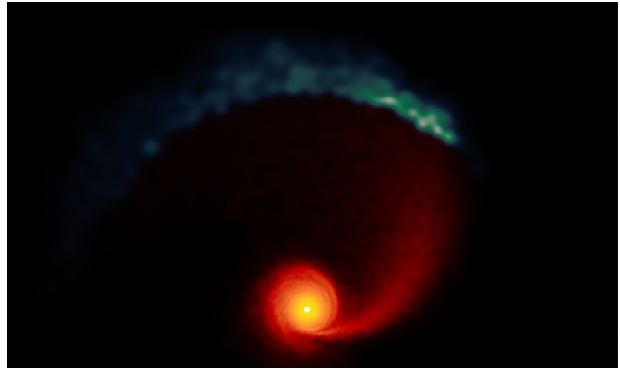}
\caption{A density slice snapshot of the 3D SNSPH simulation of a white dwarf merger of masses 0.64\msol + 1.20\msol. Material from the donor star that is no longer gravitationally bound is enhanced in brightness and colored blue. The ejecta material is represented here by 806 particles in this simulation.}
\label{fig:tail}
\end{figure}

As the secondary star disrupts, material flows through the L2 point and escapes the system. Figure \ref{fig:tail} illustrates the geometry of the simulation shortly after the complete breakup of the 0.64\msol donor star into a disk. Roughly $2.4\times10^{-3}$\msol of material achieves escape velocity ($\approx2000$ \kms), forming the end of a long tidal tail. The ejecta has an opening angle of $\approx93^\circ$ in the plane of the disk and $\approx41^\circ$ in the perpendicular direction, giving a covering factor $>25\%$ seen edge on, and $\approx9\%$ for all space. 

Simulations of other mass pairs yield very similar tidal tail ejections. Table \ref{table:masses} lists the ejecta masses from each of our SPH simulations, and as is evident from these results, tidal tail ejections massing $M_{\rm ej}\sim10^{-3}\msol$ are indeed a robust feature of double degenerate white dwarf mergers, irrespective of the white dwarf masses. Our analysis of the subsequent evolution of the tidal tails will be confined to our fiducial case of 0.64\msol + 1.20\msol.

\begin{table}[ht]
\caption{Ejected masses, $M_{\rm ej}$, from various SPH simulations of white dwarf mergers of masses $M_1$ and $M_2$.}
\centering
\begin{tabular}{ c | c | c }
\hline\hline
$M_1/\msol$ & $M_2/\msol$ & $M_{\rm ej}/\msol$\\
\hline
0.64 & 0.96 & $2.0\times10^{-3}$\\
0.64 & 1.06 & $2.3\times10^{-3}$\\
0.64 & 1.20 & $2.4\times10^{-3}$\\
0.84 & 0.96 & $1.4\times10^{-3}$\\
1.06 & 1.06 & $4.7\times10^{-3}$\\
1.06 & 1.20 & $3.3\times10^{-3}$\\
\hline
\end{tabular}
\label{table:masses}
\end{table}

Shortly after its ejection, the tidal tail expands ballistically within the nearly central potential field. It is expected that this material (at roughly 0.6 g \pcc\ in the snapshot represented in Figure \ref{fig:tail}) should drift relatively unimpeded for $\sim100$ years before the swept up ISM mass becomes non-negligible. During that time, it will expand considerably and drop in density by many orders of magnitude.   Since an SPH simulation of this expansion would be costly,  we instead evolve this material in a simple n-body code with self-gravity until it reaches homology. The merger remnant (all bound particles) is treated as a point mass for this phase of the calculation. As the SPH particles were of roughly equal mass ($m$), the density at any spatial point \textbf{r} in the n-body calculation can be approximated by $m\thinspace \bar{r}^{-3}$, where $\bar{r}$ is the average interparticle spacing around \textbf{r}. Once the material is expanding homologously, rescaling the domain is trivial.

\begin{figure}[ht]
\centering
\includegraphics[width=0.45\textwidth]{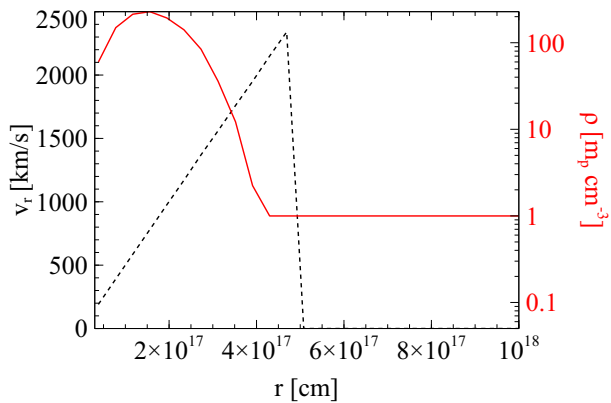}
\caption{The initial conditions of the 1D lagrangian simulation for $M_{\rm ej} = 2.7\times10^{-3}$\msol. Densities here are plotted as proton masses per cm$^{3}$ ($1.67\times10^{-24}$ g \pcc) to better illustrate the density contrast of the ejecta with the unshocked ISM. The ejected material has expanded homologously to a density of $\approx3.7\times10^{-22}$ g \pcc\thinspace and is just beginning to contact the ISM at $v_r\approx2000$ \kms.}
\label{fig:hydro0}
\end{figure}

When the material has drifted far enough to have reached a region where ISM interaction becomes relevant ($\approx5\times10^{17}$ cm), we map the resultant distribution into a 1D spherical lagrangian hydrodynamics code with a gamma-law equation of state. Figure \ref{fig:hydro0} shows the initial conditions for this phase of the calculation. Although the tails are clearly aspherical, the interaction and sweeping up of the symmetric ISM likely causes the distribution to spread and approach a more spherical configuration, in which case a 1D description is likely not too inaccurate. 

The typical density of the ISM is uncertain, however estimates of the ambient medium density around Tycho's supernova remnant by \citet{Badenes2006} suggest an ISM density of between $0.2\times10^{-24}$ g \pcc\thinspace and $5\times10^{-24}$ g \pcc, with the most probable density at $2\times10^{-24}$ g \pcc. This is consistent with roughly 1 proton mass per cm$^{3}$, and for our 1D, late-phase simulation, we use this conservative estimate, $\rho_{\rm ism}=1.67\times10^{-24}$ g \pcc.

Since radiative cooling of the shocked ejecta material is important over the long time-scales considered here, we employ an analytical cooling function of the form
\be
\frac{\Lambda}{n_en_i} = \frac{1.35\times10^{-16}}{T}+ 1.4\times10^{-27}g_{\rm ff}\sqrt{T}
\ee
for temperatures above $10^{5.5}$ K. In this expression, $n_e$ and $n_i$ are the number densities of free electrons and of the ions, respectively, and $g_{\rm ff}$ is the Kramers-Gaunt factor which can safely be set to 1.0. The second term in this expression is based on free-free emission \citep{Rybicki1986} for temperatures above $\sim10^{7.5}$, and the first term approximates the theoretical cooling curves from emission lines found in \citet{Sutherland1993} and \citet{Wiersma2009}. For simplicity we take $\Lambda/(n_e n_i) = 1.35\times10^{-21.5}$ erg~cm$^3$~s$^{-1}$ for $10^4< T <10^{5.5}$ K. This approximates the collisionally ionized equilibrium cooling rates in this range from \cite{Wiersma2009} fairly well, however sodium is expected to remain ionized at these temperatures. Since cooling rates below $10^4$ K are somewhat undetermined, and since the exact functional form of the cooling function at this low temperature does not impact the structure of the shell, we use the same constant rate down to $10^3$ K. 

Our simplified cooling function captures most of the relevant cooling physics for the shell formation with a slight decline in cooling efficiency from $10^{5.5}$ K out to $\approx10^7$ K and a steady rise in efficiency thereafter, while also being efficient enough at low temperatures to allow for neutral sodium formation. The extent to which we overestimate the neutral sodium fraction is only important insofar as our inferred optical depths will be too great, and determining precisely what the sodium ionization fraction should be is out of the scope of this work.

\begin{figure}[ht]
\centering
\includegraphics[width=0.45\textwidth]{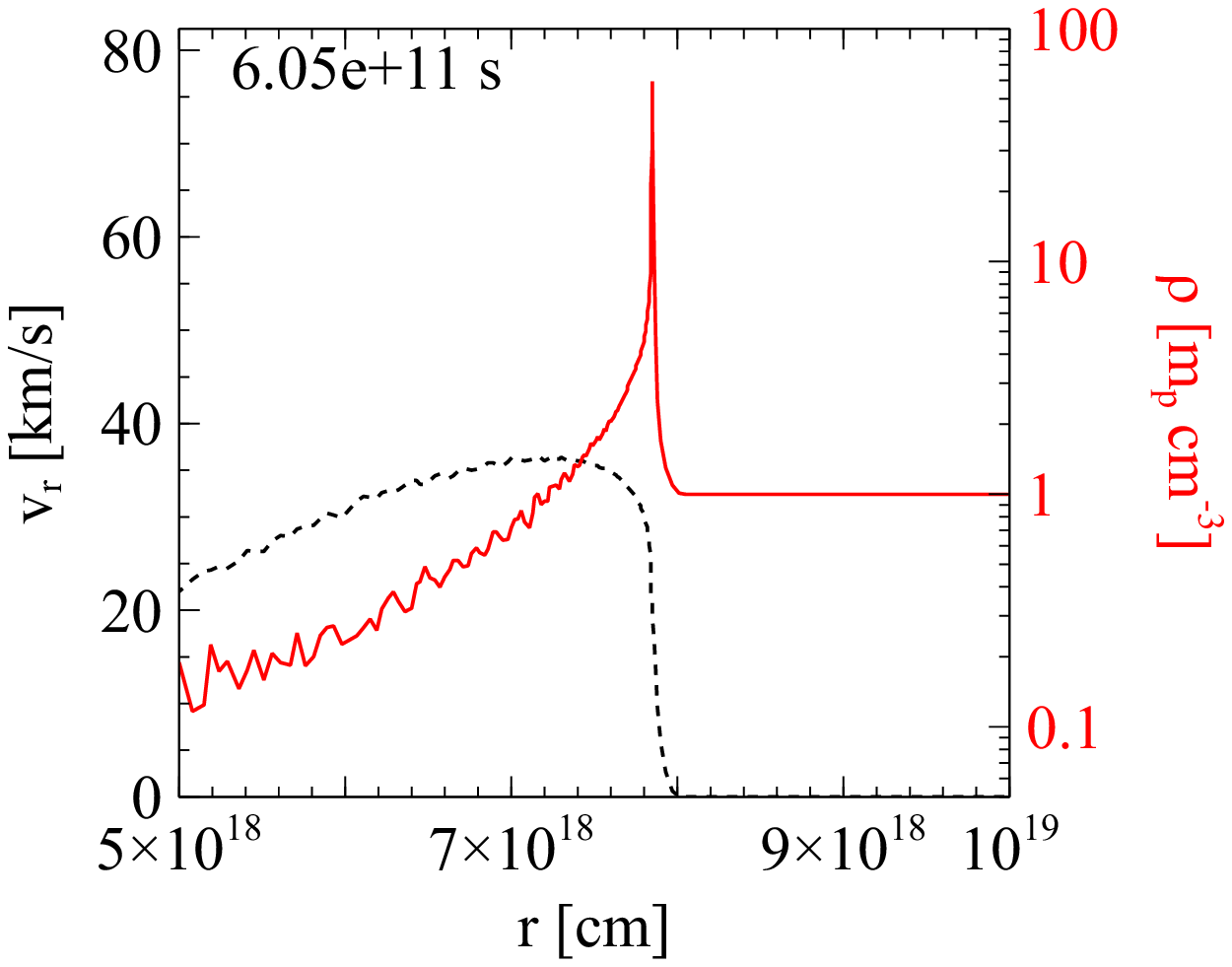}
\caption{A later stage of the 1D lagrangian simulation for $M_{\rm ej} = 2.7\times10^{-3}$\msol. The shocked ISM gas has cooled to $T<10^3$ K and formed a dense shell moving at $v_r<40$ \kms. Densities here are plotted as proton masses per cm$^{3}$ ($1.67\times10^{-24}$ g \pcc) to better illustrate the density enhancement of the shell.}
\label{fig:hydro1}
\end{figure}

When the ejecta first encounters the ISM, it is shock-heated and begins to slow as it sweeps up ISM gas. Behind the shock, the mixed ISM and ejecta material initially cannot cool, and the hydrodynamic evolution is essentially a Sedov blast wave \citep{Sedov1959}. However, as the post-shock gas expands, the temperature drops sufficiently for collisional ionization to begin to cool the gas, and after $\approx3\times10^3$ years, a thin, dense shell forms near the shock front. This shell has a mass of $\approx1.5\msol$ and a velocity $<40$ \kms. Figure \ref{fig:hydro1} shows the growth of this dense shell. 

\subsection{Analytical Comparison}

To compare these results with analytical expectations, we consider the three phases for the evolution of the expanding gas. The first is a simple ballistic phase where $r_{\rm ej}=v_{\rm ej}t$. The duration of this phase is roughly the time before the ejecta sweeps up its own mass of ISM gas, and lasts approximately 200 years. At this point, the gas will enter an energy conserving (Sedov) phase where the temperatures of the shocked shell are high and radiative losses are negligible. Evenutally, when the gas has cooled sufficiently to become radiative, it will enter a momentum conserving phase.

During the energy conserving phase of the interaction, when the adiabatic evolution is that of a Sedov blast wave \citep{Sedov1959, Chevalier1982}, the location of the contact discontinuity is, for a uniform ISM, roughly of the form
\be
r_c\approx\left(\frac{E_{\rm ej}t^2}{\rho_{\rm ism}}\right)^{1/5}\approx0.87\left(\frac{M_{\rm ej}v_{\rm ej}^2 t^2}{\rho_{\rm ism}}\right)^{1/5},
\ee
where $E_{\rm ej}$ is the kinetic energy of the ejecta material. The shock velocity then evolves as
\be
v_s\approx0.87 M_{\rm ej}^{1/5} v_{\rm ej}^{2/5} \rho_{\rm ism}^{-1/5} t^{-3/5}. 
\ee
The jump conditions for a strong shock give a post-shock pressure of the form
\begin{equation}
P  = \frac{1}{2} \frac{\rho_{\rm ism} v_{\rm s}^2 }{\gamma  + 1}=2n_{\rm ism}kT_s.
\end{equation}
Here, $\gamma$ is the usual adiabatic index, and the post-shock temperature for an ideal gas is then
\begin{equation}
T_s \approx1.2\times10^7\mu v_{s,8}^{2}~{\rm K},
\end{equation}
where $v_{s,8} = v_s/10^8~\cms$. 
The timescale for this post-shock gas to cool is
\begin{equation}
t_{\rm cool} = \frac{n k T_s}{(\gamma-1)\Lambda}.
\end{equation}
Employing an approximation for the cooling function which applies in the range $10^5$ K$< T < 10^{7.3}$ K from \citet{Draine2011}, $\Lambda/(n_e n_{\rm H})\approx1.1\times10^{-22}T_6^{-0.7}$ erg cm$^3$ s$^{-1}$, and combining the previous relations gives for a nominal ISM number density of 1 \pcc
\ba
\nonumber t_{\rm cool}&\approx&120~T^{1.7}~{\rm s}\\
&\approx&2.3\times10^{34} M_{-3}^{0.68} v_{\rm ej,8}^{1.36} t^{-2.04}~{\rm s},
\ea
where $M_{-3}=M_{\rm ej}/10^{-3}\msol$ and $v_{\rm ej,8}=v_{\rm ej}/1000$ \kms. The time to cool ($t_{\rm cool}\sim t$) is then roughly $10^4$ years.

At this stage, the gas becomes radiative and energy is no longer conserved. The shell will reach a velocity $v_f$ given by conservation of momentum 
\begin{equation}\label{eq:momentum}
M_{\rm ej} v_{\rm ej} = M_{\rm sh} v_f,
\end{equation}
where $M_{\rm sh}$ is the final mass of the shell, which for the case $v_f \ll v_{\rm ej}$ will be dominated by swept up ISM gas. Assuming the ISM is of constant number density, $n_{\rm ism}$, the radius at which the shell slows to a value $v_f$ is then
\ba\label{eq:rmomentum}
r_{\rm sh} &=& \left( \frac{3}{4 \pi} \frac{ M_{\rm ej}}{\mu m_p n_{\rm ism}} \frac{v_{\rm ej}}{v_f}
\right)^{1/3} 
\\
&\approx& M_{-3}^{1/3} v_{\rm ej,8}^{1/3} v_{f,6}^{-1/3} \mu^{-1/3} {n^*_{\rm ism}}^{-1/3}~ {\rm pc},\nonumber
\ea
where $v_{f,6}=v_f/10$ \kms and $n^*_{\rm ism}$ is the dimensionless quantity $n_{\rm ism}/$\pcc. For our purposes here, these quantities are all of order unity, and the predicted location of the shell is $\approx1.1$ pc. Taking the time derivative of equation (\ref{eq:rmomentum}) gives the drift time
\be
t_{\rm drift} = \frac{1}{4}\left( \frac{3}{4 \pi} \frac{ M_{\rm ej}v_{\rm ej}}{\mu m_p n_{\rm ism}} \right)^{1/3}
\left[v^{-4/3}\right]_{v_i}^{v_f},
\ee 
where $v_i$ is the velocity at the beginning of the radiative phase. This drift time is also $\sim10^4$ years, and so the total time from the ejection of the tidal tail to the interaction with the ISM and subsequent slowing to a cooled, radiative shell moving at $\approx40$ \kms\ is $\approx2\times10^4$ years. This compares well with the simulated result of $1.9\times10^4$ years.

%Shen \etal (2012)  have demonstrated that the thermal timescale for heat diffusion into the merged remnant core is of order $\sim10^4$ years. If this mechanism is responsible for igniting a SNIa, this timescale would be comparable to the drift and cooling time of the absorber shell. 

\begin{figure}[ht]
\centering
\includegraphics[width=0.45\textwidth]{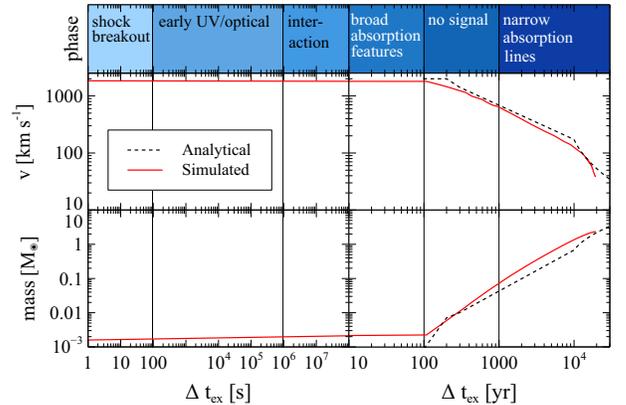}
\caption{The evolution of the ejecta mass interacting with the ISM, forming an absorbing shell. Both the results of a 1D lagrangian simulation and the analytical estimates at $\Delta t_{\rm ex}>100$ yr are shown here. The phases illustrated here are separate from those discussed in \S2 and are discussed in more detail in \S3.}
\label{fig:evo}
\end{figure}

Figure \ref{fig:evo} diagrams the evolution of the shock through the three phases outlined here, comparing the analytical estimates to the simulation results. The analytical estimates are only relevant after an initial drift period, during which the ejecta mass is much greater than the swept up ISM mass. The shock radius during the energy conserving phase evolves as a 2/5ths power-law, and during the cooling phase, as a 1/4th power law. At this stage, the shell velocity in the simulation becomes difficult to measure accurately as without clumping and self-gravity, it begins to expand adiabatically into the surrounding medium.

\section{Observational Signatures}

The potentially detectable signatures of the tidal tails depend primarily on the lag time, \dtx, between when the tails are ejected and when the merged system explodes as a SNe~Ia (assuming that it does explode).  The value of \dtx\ is unclear given our incomplete understanding of the post-merger evolution, but obviously interesting physical timescales include the dynamical ($\dtx \sim 10^2-10^3$~secs), the viscous ($\dtx \sim 10^4-10^8$~secs), and the thermal ($\sim 10^4-10^5$~years) times.  For resonantly induced collisions \citep{Katz2012} a relevant timescale is the period of close approaches ($\sim 10^4$~years). Without adopting any specific theoretical paradigm, we consider the possible observable signatures as a function of \dtx.  

\subsection{$\dtx \lesssim 100$~s: shock breakout and cooling}

If the system detonates promptly, on the dynamical timescale, the tidal tails will have only had time to reach radii of $r \lesssim 10^{10}$~cm before they are impacted by the supernova ejecta.   At these times, the CSM is optically thick, and the interaction will drive a radiative shock through it.  Emergence of this shock should produce a brief  x-ray pulse similar to shock breakout in core collapse supernovae \citep{Chevalier1979, Imshennik1981, Hoflich2009}.  Based on scaling relations derived for x-ray bursts in SNe~Ia \citep{Piro2010, Kasen2010} we anticipate  bolometric luminosities of order $10^{44}~{\rm ergs~s^{-1}}$, peaking in the $1-10$~keV range and lasting roughly a light crossing time, $t \sim 10^{-2} \dtx \sim 1$~sec. 

Subsequent cooling of the shock heated layers will continue to produce UV/optical emission for some time after breakout, although this will be relatively dim and difficult to detect unless the supernovae is caught extremely early.  The cooling emission will also be anisotropic, and may only be prominent for viewing angles of the half of the system where the tidal tail is ejected.  The early observations of SN~2011fe \citep{Nugent2011, Bloom2012} limit the emitting radius to be $\lesssim 1.5 \times 10^9$~cm, which corresponds to a lag time $\dtx \approx 10$~s.    This constraint assumes that the explosion time has been accurately determined by extrapolating the early light curve, which is 
not trivial \citep{Piro2012}.  More detailed modeling is  required, but it is likely that the early observations of SN~2011fe and other SNe~Ia (e.g., SN~2009gi, Foley \etal 2012) are sensitive to tidal tail interaction and hence constrain WD detonations from violent mergers.

\subsection{$100~{\rm s} \lesssim  \dtx \lesssim 10^6$~s: early UV/optical emission}

For this range of lag times  (comparable to the viscous timescale of the remnant disk) the CSM is confined to radii $r \lesssim 10^{13}$~cm and will be impacted by the supernova ejecta within a few hours after the explosion.   The resulting shock breakout burst will be longer lasting ($\sim$minutes) and fairly soft ($\sim 100$~eV).  Following this,  optical/UV cooling emission from the expanding, shock-heated layers should be quite bright, $L_{\rm bol} \approx 10^{42}-10^{43}$ erg s$^{-1}$. Statistical studies of early SN~Ia light curves have strongly ruled out  shock emission of this magnitude in the first few days after explosions \citep{Bianco2011, Hayden2010}.  However, because the total mass of the tidal tails is relatively small, the duration of that emission should be fairly short -- of order the diffusion time through the shocked region, or about half a day  \citep{Piro2012}.  Emission this early may have been missed in the statistical studies.    For the case of  SN~2011fe \citep{Nugent2011, Bloom2012}, the early light curve data clearly rules out tidal tail interaction at these radii, unless one believes that the explosion time had been underestimated by $\sim 0.5$~days.

\subsection{$10^6~{\rm s} \lesssim  \dtx \lesssim 10^8$~s: interaction}

For these lag times, the CSM is spread over radii $r \lesssim 10^{14}-10^{16}$~cm, and the supernova ejecta may experience interaction over timescales of weeks. One expects  a range of  signatures, in particular the shocked gas should produce radio and x-ray radiation due to synchrotron and inverse compton scattering, while the photoionization of the unshocked CSM may produce optical emission lines.  Relatively
speaking, the density of the tidal tail CSM at these phases is  high, with an effective mass loss rate in the range  $\dot{M} \approx M_{\rm ej} (v_{\rm ej}/ v_{\rm w})/\dtx \approx 10^{-2} - 10^{-5} ~\msol~{\rm yr}^{-1}$ depending on the tidal tail mass and the wind velocity, $v_{\rm w}$.  This is significantly greater than the CSM density of most companion star winds in the single degenerate scenario.

A variety of SN~Ia observations at x-ray/optical/radio wavelengths  have constrained the mass loss rate to be $\lesssim 10^{-6} - 10^{-5}~\msol~{\rm yr}^{-1}$ for several individual events \citep{Hughes2007, Mattila2005, Cumming1996, Panagia2006, Russell2012}.  The observations of SN~2011fe are the most constraining, limiting $\dot{M} \lesssim 10^{-8} ~\msol~{\rm yr}^{-1}$ \citep{Horesh2012, Margutti2012, Chomiuk2012}.  These results apparently exclude white dwarf merger models with lag times in this range.

\subsection{$10^8~{\rm s} \lesssim  \dtx \lesssim 100$~yrs: ''broad'' absorption features}

Over these lag times, the bulk of the CSM has expanded ballistically to $r \sim 10^{17}$~cm and it will take years for the supernova ejecta to interact with it.  However, one may still be able to see the tidal tails in absorption. The tails will have cooled adiabatically and are likely neutral with NaI columns of order $10^{13} - 10^{15}~{\rm cm^{-2}}$.  The NaID lines will then be optically thick and will produce absorptions for those viewing angles where we are looking through the tails, or about 10\% of the time. The width of these CSM absorptions will be of order the escape velocity, $\sim 10^8~\cms$, which is  narrower than typical supernova lines, but much broader than typical CSM absorptions.  There is therefore some hope of uniquely identifying lines associated with tidal tail ejections.  NaID absorption features of this width have not been identified in any SNe~Ia, despite the enormous number of optical spectra obtained. This would apparently rule out white dwarf merger remnants that explode with these lag time scales.

\subsection{$100~{\rm yrs} \lesssim \dtx  \lesssim 10^3$~yrs: no signature}

After roughly 100 years, the expanding tails have swept up enough ISM material that they have been
entirely shocked and are in the Sedov phase.  The hot shell of gas will be ionized, and  optically thin to all relevant lines.  One therefore does not expect any prominent observational signatures again  
until the shell has begun to radiatively cool.

\subsection{$10^3~{\rm yrs} \lesssim \dtx  \lesssim 10^5$~yrs: narrow  absorption lines}

After $10^3$~yrs, the tidal tail ejecta have slowed to the point that radiative cooling becomes important and the CSM forms a  dense shell moving at velocities $\lesssim 100~\kms$.  The interaction with the CSM likely causes the tails to spread and take on a more spherical shell configuration.  Absorption lines may then be seen from the cool shell, at least in a statistical way.  We can calculate the optical depth of the NaID line (at line center) through a shell of density $n_{\rm sh}$  and thickness  $\Delta r$ as
\begin{equation}\label{eq:tau}
\tau = n_{\rm sh} x_{\rm NaI} \sigma_0 \Delta r,
\end{equation}
where $x_{\rm NaI}$ is the fraction of sodium in the neutral state and $\sigma_0$ is the cross-section, per atom, of the NaID line at line center.  Assuming that the width of the line is set by Doppler broadening at a velocity
$\Delta v \sim 10~\kms$, this cross-section is
\begin{equation}\label{eq:sigma}
\sigma_0 = \frac{\pi e^2}{m_e c}  \frac{\lambda_0}{\Delta v} f_{\rm osc} x_{\rm NaI} A_{\rm Na}
= 10^{-18}~\Delta v_{6}^{-1} x_{\rm NaI} ~{\rm cm^2},
\end{equation}
where we take the number abundance of sodium to be solar, $A_{\rm Na} = 2\times10^{-6}$. The neutral fraction of sodium, $x_{\rm NaI}$ is difficult to estimate, since it likely requires cooling below the recombination temperature of hydrogen.  In the absence of detailed calculations, we assume here that the neutral fraction is of order unity.  

For a thin shell of thickness $\Delta r$, the shell density is $\rho_{\rm  sh} = M_{\rm sh}/(4\pi r_{\rm sh}^2 \Delta r)$  and the line optical depth  is  independent of the shell thickness.  Using equations (\ref{eq:tau}), (\ref{eq:sigma}), and (\ref{eq:rmomentum}), and given that the swept up ISM mass dominates the mass of the shell, we can then write the optical depth of the NaI lines in terms of the density of the ISM and the final velocity of the shell as
\ba
\nonumber\tau &=& \left(\frac{16\pi^2 M_{\rm ej}n_{\rm ism}^2  (v_{\rm ej}/v_f)}{9\mu m_p}\right)^{1/3}\times10^{-18}\Delta v_{6}^{-1} x_{\rm NaI}\\
&\approx& 1.39~  x_{\rm NaI} \left(\frac{{n^*_{\rm ism}}^2}{v_{f,6}}\right)^{1/3}.
\ea
For the values used in this paper ($n_{\rm ism}=1$ and $v_{f,6}=4$), this reduces to $\tau\approx0.87$. That the optical depth depends only on the ISM density and the final velocity of the shell should not be too surprising as an undisturbed ISM will reach an optical depth of $\sim1$ for sodium at a radius where $v_{f,6}\approx5$. Without the moving shell, the Na absorption would appear as narrow ISM lines (not blue-shifted), as is seen for other transient types. Put another way, if the source of the blue-shifted absorption were from a shell of shock-cooled ISM gas (as is the case here since $M_{\rm ej}/M_{\rm sh}<<1$), requiring an optical depth $\sim1$ sets a radius (similar to the same requirement for unshocked gas, modulo a factor of 3) at which $v_{f,6}\approx5$ naturally, provided $M_{\rm ej}$ is sufficiently large for equation (\ref{eq:rmomentum}) to hold.

\begin{figure}[ht]
\centering
\includegraphics[width=0.45\textwidth]{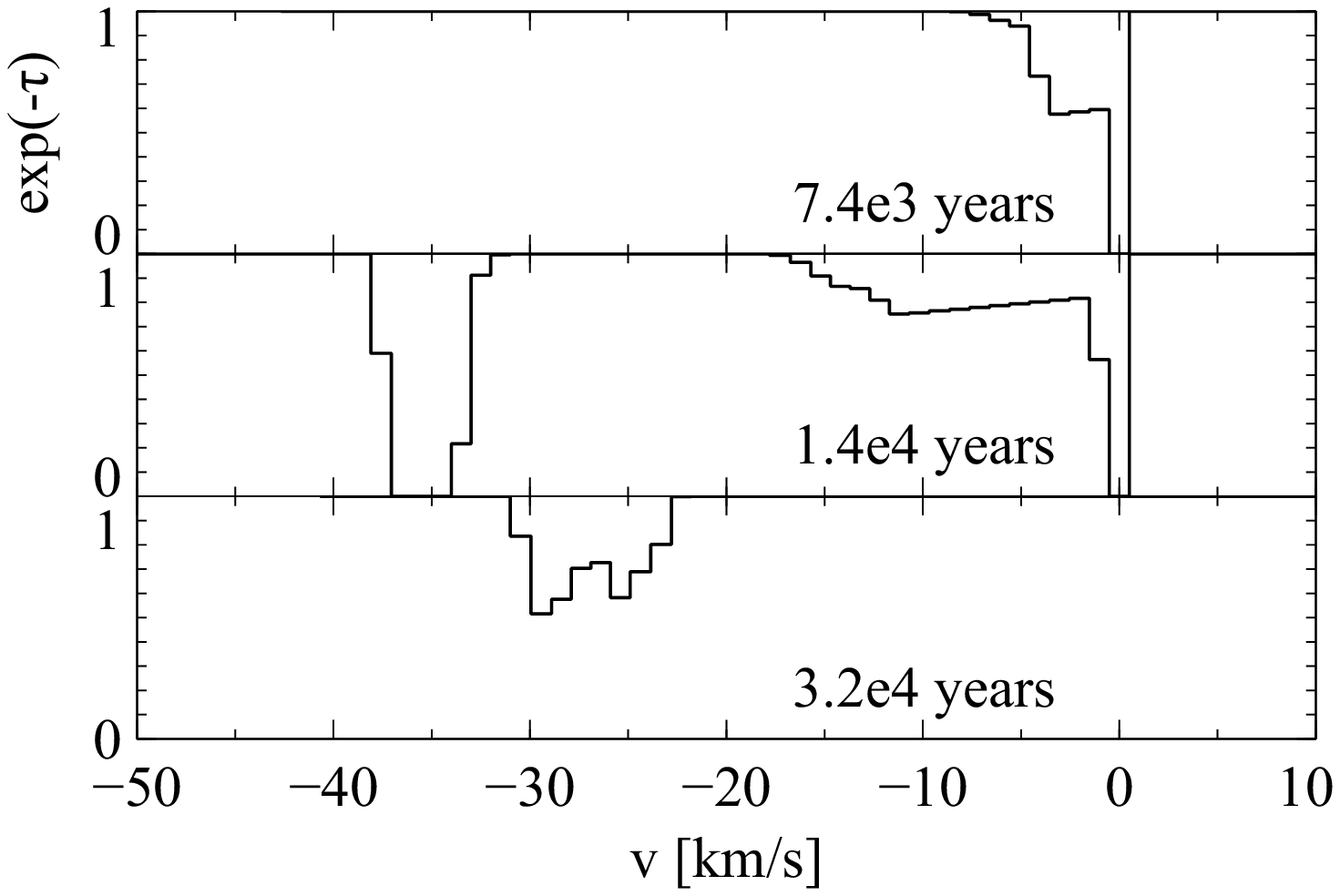}
\caption{Sodium absorption profiles for a range of times from the early phase of the formation of the shell, when the shell has not had sufficient time to cool in order for neutral sodium to form, to the narrow absorption line phase, to the late-time disruption of the shell due to adiabatic expansion. At very late times, the zero velocity absorption material has all been swept up in the finite-sized simulation box.}
\label{fig:lines}
\end{figure}

In order to synthesize an absorption line profile for our simulations, we simply integrate equation (\ref{eq:tau}) numerically for every wavelength $\lambda$ and use a gaussian dispersion about the line center $\lambda_0$ for thermal broadening, $\Delta\lambda = (\lambda_0/c)(\sqrt{kT/\mu m_p})$. We take $x_{\rm NaI}$, the neutral fraction of sodium, to be 1.0 everywhere where $T<5\times10^3$ K. Figure \ref{fig:lines} shows the evolution of the neutral sodium absorption line through various phases of our simulation. 

In a simple model, such as this, the shell cannot persist for long as it is at a higher pressure than its surroundings. Without clumping due to self-gravity, the shell simply begins to expand into the surrounding medium and the subsequent drop in density precludes the synthesis of any strong absorption lines at much later times. It is expected that clumping from Raleigh-Taylor instabilities not captured in a 1D simulation should also enhance absorption at later times. These factors work to constrain the available time in our simulation for sodium absorption. However, we expect a multi-D approach incorporating self-gravity should, for the most part, validate our results and extend the available absorption time. If we compare our synthesized line profiles from times before adiabatic expansion to observed absorption lines for SN~2008fp in \citet{Sternberg2011}, we find a good correspondence for the absorption profile of the blue-shifted sodium lines.

\section{Discussion}

We have explored the observational consequences of the tidal tails believed to be ejected in SNe~Ia resulting by white dwarf mergers.  The tails produce several observational signatures that may constrain the lag time between the merger and the subsequent explosion of the remnant (assuming an explosion does occur).

If the lag time is very short (\dtx$\lesssim100$ s), the interaction of the supernova ejecta with the tidal tail ejecta should produce brief x-ray pulses like those observed during the shock breakout phase of core collapse supernovae. If the lag time is slightly longer (\dtx$ \sim 10^4$ s), the interaction shock  results in early ($\sim$day) UV and optical emission. For lag times $\gtrsim 10^8$ s, the tidal tail ejecta material is  too distant to be impacted by the supernova ejecta in a meaningfully short time, but the tail material may still be seen
in absorption features that are broader than typical CSM absorption lines.  The lack of observed shock emission or broader CSM absorption features in the SN~Ia sample would appear to exclude mergers exploding with lag times in the range $10^4~{\rm s} \lesssim \dtx \lesssim 100$~yr, which places important constraints on models presumed to explode on a viscous timescale.

By combining the results of a 3D hydrodynamics simulation with simple 1D models, we have shown that given time to interact with the ISM (\dtx$\gtrsim100$ yr), these mass ejections produce Sedov shock waves in the surrounding medium which then evolve into thin, over-dense shells after radiatively cooling (\dtx$\gtrsim10^4$ yr). The covering factor for the ejecta is $<10\%$, however, interaction with the ISM will likely spread the tail material and increase the covering fraction to something closer to $\sim 50\%$.  Multi-dimensional models will be needed to determine the geometry and the sub-structure that develops due to hydrodynamical instabilities. 

Depending on the exact value of \dtx, the resulting shock front can have a range of velocities from 10-100 \kms, reproducing the observed range of blue-shifted sodium absorption lines in SNeIa. The timescale for the evolution of this absorption material into a shell is also comparable  to the timescale for the thermal evolution or helium shell burning time of white dwarf merger remnants, $\sim10^4-10^5$ years.  The optical depths of these shells are of order $\sim 1$, independent of many of the simulation parameters. Synthesizing absorption lines from our hydrodynamical simulations yields absorption profiles  similar to those found in \citet{Sternberg2011}. 

Although we have focused here on CSM due to tidal tail ejection, there may be other means of creating mass outflow in merging double-degenerate systems.  These include mass outflows during the rapid accretion that immediately precedes the merger \citep{Dan2011, Guillochon2010}, disk winds arising during the viscous evolution \citep{Ji2013}, or mass ejections in the post-main-sequence evolution of the white dwarf progenitors
\citep{Shen2013}. Another interesting case is the possible tidal stripping that may occur during close encounters in triple systems, driven by Kozai resonances \citep{Katz2012}. While \citet{Katz2012}  only consider ``clean case" scenarios, there is potentially a large parameter space available for Kozai resonances to induce tidal stripping at the last closest encounter before the two of the stars collide.   Further simulation will be needed to quantify the mass loss by some of these mechanisms.   Tidal tail ejection, however, appears to be a robust prediction of white dwarf mergers; searching for the resulting signatures should therefore provide interesting constraints on the progenitors of SNe~Ia.

\section*{Acknowledgments}
We thank Ken Shen and Ryan Foley for their input on NaID line observations of SNe~Ia, and William Gray and Evan Scannapieco for their input on methods to simulate the radiative cooling phase. We also thank our anonymous referee for their insightful comments and suggestions to improve this manuscript.
This research was supported by an NSF Astronomy and Astrophysics Grant (AST-1109896) and by 
the DOE SciDAC Program (DE-FC02-06ER41438) and by the Director, Office of Energy
Research, Office of High Energy and Nuclear Physics, Divisions of
Nuclear Physics, of the U.S. Department of Energy under Contract No.
DE-AC02-05CH11231.
 This research used resources of the National Energy Research Scientific Computing Center, which is supported by the Office of Science of the U.S. Department of Energy under Contract No. DE-AC02-05CH11231. 
We are grateful for computer time supplied by the Advanced Computing Center at Arizona State University.

\bibliographystyle{apj}

\end{document}